\documentclass[12pt]{article}

\newcounter{fig}

\newcommand{\del}{\partial}
\newcommand{\delk}{\partial_{k_z}}
\newcommand{\kdag}{\mathbin{k\mkern-10mu\big/}}
\newcommand{\deldag}{\mathbin{\partial\mkern-10.5mu\big/}}
\def\tgs {{\tilde g^s}}
\def\gvp {{\gamma_\|}}
\def\vgva{{v_1 \gamma_\|}}
\def\vgvb{{v_2 \gamma_\|}}

\newcommand{\beq}{\begin{equation}}
\newcommand{\eeq}{\end{equation}}
\newcommand{\beqa}{\begin{eqnarray}}
\newcommand{\eeqa}{\end{eqnarray}}

\newcommand{\cpdag}{\mathbin{{\rm CP}\mkern-20.0mu\big/}}
\def\scp{s_{\cpdag}}

\begin{document}

%%%%%%%%%%%%%%%%%%%%%%%%%%%%%%%%%%%%%%%%%%%%%%%%%%%%%%%%%%%%
%
% title page
%
%%%%%%%%%%%%%%%%%%%%%%%%%%%%%%%%%%%%%%%%%%%%%%%%%%%%%%%%%%%%
\begin{titlepage}
\pagestyle{empty}

\rightline{CERN-TH/2002-011}
\rightline{NORDITA 2002-5 HE}
\rightline{HD-THEP-02-7}
%\rightline{February 2002}
 \rightline{\today}
\vskip 0.4in

\begin{center}
     {\bf\Large Semiclassical force for electroweak baryogenesis:} 
\vskip 0.1in
     {\bf\Large three-dimensional derivation}
\end{center}

\vskip 0.3in

\begin{center}
  {\large Kimmo Kainulainen$^\bullet$, Tomislav Prokopec$^\diamond$,}\\
\vskip 0.05in
  {\large Michael G. Schmidt$^\diamond$ and Steffen Weinstock$^\diamond$} \\
\vskip 0.3in
  {\it $^\bullet$Theory Division, CERN, CH-1211, Geneva, Switzerland 
  and \\ NORDITA, Blegdamsvej 17, DK-2100, Copenhagen \O , Denmark} \\
  \vskip 0.1in
  {\it $^\diamond$Institute for Theoretical Physics, Heidelberg University\\
           Philosophenweg 16, D-69120 Heidelberg, Germany} \\
\end{center}

\vskip 0.5in

% abstract:

\centerline{\bf Abstract}
\vskip 0.3truecm

\noindent
We derive a semiclassical transport equation for fermions
propagating in the presence of a CP-violating planar bubble wall at a
first order electroweak phase transition.
Starting from the Kadanoff-Baym (KB) equation for the two-point (Wightman) 
function we perform an expansion in gradients, or equivalently 
in the Planck constant $\hbar$. We show
that to first order in $\hbar$ the KB equations have a spectral solution,
which allows for an on-shell description of the plasma excitations. 
The CP-violating force acting on these excitations is found to be
enhanced by a boost factor in comparison with the 1+1-dimensional
case studied in a former paper.
We find that an identical semiclassical force can be obtained by
the WKB method. Applications to the MSSM are also mentioned.

\vskip 0.7in

\footnoterule
\vskip 3truemm
{\small\tt
\noindent$^\bullet$kimmo.kainulainen@cern.ch \\ {\rm University of 
Jyv\"askyl\"a, Finland, on leave of absence.}\\
$^\diamond$T.Prokopec@, M.G.Schmidt@, S.Weinstock@thphys.uni-heidelberg.de}

\end{titlepage}

\baselineskip=18pt

%%%%%%%%%%%%%%%%%%%%%%%%%%%%%%%%%%%%%%%%%%%%%%%%%%%%%%%%%%%%
%
% main text
%
%%%%%%%%%%%%%%%%%%%%%%%%%%%%%%%%%%%%%%%%%%%%%%%%%%%%%%%%%%%%
%
%%%%%%%%%%%%%%%%%%%%%%%%%%%%%%%%%%%%%%%%%%%%%%%%%%%%%%%%%%%%
%
% INTRODUCTION
%
%%%%%%%%%%%%%%%%%%%%%%%%%%%%%%%%%%%%%%%%%%%%%%%%%%%%%%%%%%%%
%
\section{Introduction}
\label{Introduction}

Modeling electroweak baryogenesis
(EWBG)~\cite{KuzminRubakovShaposhnikov}
requires a study of the generation and transport of CP-violating 
flows arising from interactions of fermions with the expanding phase 
transition fronts. Indeed, the most prominent problem in the EWBG
scenario over the past few years has been to find a systematic derivation
of the appropriate transport equations~\cite{CohenKaplanNelson:1991}, 
including the CP-violating sources. This paper is the second in a 
series dedicated to reach this goal. 
In the first paper~\cite{KPSW1} (KPSW-I) we have derived
a semiclassical transport equation starting with the Kadanoff-Baym (KB)
equations~\cite{SchwingerKeldysh,KadanoffBaym:1962}
for the Wightman out-of-equilibrium two-point function.
For simplicity we have restricted ourselves to 1+1 dimensions there.
% and we have considered only the collisionless limit.
%In this case the relevant KB equations 
%can be easily obtained from the Dirac equation. 
An important tool for the derivation was an expansion in gradients 
in the center-of-mass coordinate. This is a good 
approximation when the typical de Broglie wave length $\ell_{\rm dB} 
\sim 1/T$ of a particle in the plasma is small in comparison to 
the wall thickness, $\ell_{\rm dB} \ll \ell_{\rm wall}$. This is the case
for example in many supersymmetric models, which typically have
$\ell_{\rm wall} \sim (10-20)/T$~\cite{ThickWalls}. 
Here we extend the derivation to the more realistic case of fermions 
interacting with a planar phase transition front in 3+1 dimensions.  
As in~\cite{KPSW1} we only consider collisionless fermions 
and we do not account for the gauge degree of freedom. 
Some aspects of collision terms were discussed 
in~\cite{Kainulainen:2002sw} and the gauge fields were studied in the
1+1 dimensional case in~\cite{Kainulainen:2002ah}.

A crucial step in our derivation is to show that, to second order 
in gradients or equivalently to first order in $\hbar$, the KB equation
for the Wightman function $G^<$ written in 
the mixed (Wigner) coordinates admits a {\em spectral solution}. This 
is not a generic feature, and does not persist beyond the order $\hbar$.  
Nevertheless, to first order in $\hbar$ one can write an effective
kinetic equation for a suitably chosen distribution function for the 
corresponding on-shell excitations. 
This kinetic equation has the familiar Liouville form, but
with a semiclassical force term that includes a CP-violating part
of order $\hbar$ which sources baryogenesis. This CP-violating force
is in agreement with the earlier results~\cite{ClineJoyceKainulainen-II}
derived by use of WKB-methods~\cite{JPT-letter,JPT-thick,HuberJohnSchmidt}.

It should be stressed that the existence of the spectral 
solution to order $\hbar$ is a new result. Indeed, while fermion 
dynamics in the presence of classical background fields have been 
studied extensively within the Schwinger-Keldysh 
formalism~\cite{ElzeGyulassyVasak:1986,
Bialynicki-BirulaGornickiRafelski:1991,ZhuangHeinz-I,ShinRafelski,
ZhuangHeinz-II}, apparently no attempts have been made to find explicit 
solutions beyond the classical approximation. 
Neither the fact that the spectral decomposition ansatz fails 
beyond the order $\hbar$ seems to be known. Our result is also crucial for 
EWBG-calculations since no CP-violating terms appear in the 
classical mean field approximation. Moreover, as the failure of 
the spectral decomposition ansatz beyond order $\hbar$ means that 
higher order effects cannot be described in the semiclassical 
limit, we are very lucky that in the most interesting cases for 
the EWBG the walls are thick enough so that the  gradient expansion 
converges rapidly and our semiclassical theory can be expected to 
be a very good approximation. 

This paper is organized as follows. In section 2 we derive the 
Kadanoff-Baym equation for a collisionless fermionic system
and introduce the Lorentz-boost that relates the wall 
rest frame and the local frame with no parallel momentum, where the
KB equations can be reduced to the 1+1-dimensional form. We then
construct an explicit connection between the two frames and 
make use of our old 1+1 dimensional results~\cite{KPSW1} to derive 
the semiclassical transport equations appropriate for the EWBG-problem
in 3+1 dimensions. In section 3 we consider the transformation properties 
of fermionic currents and relate the CP-violation in the fluid 
equations to the non-conservation of the axial current divergence. 
In section 4 we derive the  semiclassical force in the WKB-approach 
and in section 5 we apply our results to chargino transport in the 
MSSM.  Finally, section 6 contains our conclusions.
% and finally, in the appendix 
%we consider relaxing the requirement of strict static planar symmetry 
%in the wall frame.

%
%%%%%%%%%%%%%%%%%%%%%%%%%%%%%%%%%%%%%%%%%%%%%%%%%%%%%%%%%%%%
%
% DERIVATION OF KINETIC EQUATIONS
%
%%%%%%%%%%%%%%%%%%%%%%%%%%%%%%%%%%%%%%%%%%%%%%%%%%%%%%%%%%%%
%
\section{Derivation of kinetic equations}
\label{Derivation of kinetic equations}

As in KPSW-I~\cite{KPSW1}, we will consider a fermionic field 
described by the Lagrangian
\beq
     {\cal L} = i\bar{\psi}\deldag\psi - \bar{\psi}_Lm\psi_R
           - \bar{\psi}_Rm^*\psi_L + {\cal L}_{\rm int},
\label{lagrangian}
\eeq
where the mass
\beq
     m(u) = m_R(u) + i m_I(u) = |m(u)|\mbox{e}^{i\theta(u)}
\label{mass1}
\eeq
arises from an interaction with a spatially varying scalar field 
condensate. This situation is realised at the first order electroweak 
phase transition (EWPT), which proceeds via nucleation and growth of 
the broken 
phase bubbles of a nonzero Higgs condensate. EWPT bubbles quickly grow 
very large in comparison with the thickness of the phase transition 
front, so one can to a good approximation assume a planar symmetry 
on the scale of the bubble walls.  As a result we can take $m$ 
to be a function of one spatial coordinate. Moreover, bubble walls 
are often found to be very wide~\cite{ThickWalls} in comparison 
with the typical de Broglie wavelength of a particle in the plasma, 
$\ell_{\rm wall} \ll \ell_{\rm dB} \sim 1/T$. This suggests that one 
should be able to treat the kinetics of plasma excitations near the 
phase transition front in an expansion in gradients. 

Here we consider only a collisionless plasma. Interactions are 
specified by the Lagrangian ${\cal L}_{\rm int}$, and in addition to 
reproducing the usual collision terms in kinetic theory, they give 
rise to a new CP-violating source~\cite{Kainulainen:2002ah}. We will
study these effects in detail elsewhere~\cite{KPSW3}. For simplicity 
we also restrict ourselves to a case without gauge degrees of freedom. 
Transport equations with a self-consistent electric field were 
considered in 1+1-dimensions in Ref.~\cite{Kainulainen:2002ah}. \\

As in~\cite{KPSW1}, the basic object of our study is the dynamical two point 
(Wightman) function
\beq
   G^<_{\alpha\beta}(u,v) 
    =  i\left< \bar{\psi}_\beta(v)\psi_\alpha(u) \right>,
\label{G_less}
\eeq
where the expectation value is taken over the initial state. In the 
collisionless case the equation of motion for $G^<$ can be obtained
by use of the Dirac equation that follows from the Lagrangian 
(\ref{lagrangian}):
\beq
    \Big( i\deldag_u - m_R(u) + im_I(u)\gamma_5 \Big) \psi(u) = 0.
\label{Diraceqn}
\eeq
In order to establish the gradient expansion, we first separate the 
average center-of-mass coordinate $x=(u+v)/2$ from the relative coordinate 
$r=u-v$ by performing the Wigner transform (Fourier transform with respect 
to $r$) to the mixed representation:
\beq
    G^<(x,k) = \int d^{\,4} r \, e^{ik\cdot r} G^<(x+r/2, x-r/2).
\label{wigner1}
\eeq
For further reference we note that $G^<(x,k)$ satisfies the hermiticity 
property
\beq
    \left[ i\gamma^0G^<\right]^\dagger =  i\gamma^0G^<
\label{G_herm}
\eeq
implied by the definition~(\ref{G_less}). Using (\ref{Diraceqn}) and 
transforming to the mixed representation, one finds the equation of 
motion
\beq
    \left( \kdag  +  \frac{i}{2}\deldag 
     - m_R(x-\frac{i}{2}\partial_k) 
     - i\gamma^5 m_I(x-\frac{i}{2}\partial_k) \right)
   iG^< = 0 \,,
\label{G_less_eom}
\eeq
where
\beq
  m_{R,I}(x-\frac{i}{2}\partial_k) \equiv m_{R,I}(x) 
   \mbox{e}^{-\frac{i}{2}\stackrel{\leftarrow}{\del}\!\cdot\,\del_{k}}.
\label{mass}
\eeq   
Formally equation (\ref{G_less_eom}) is completely general and 
valid up to any order in gradients and for arbitrary space and time 
dependence of the mass term (\ref{mass}). This expansion in gradients 
of the center-of-mass coordinate $x$ can be also viewed as an expansion
in powers of the Planck constant. We have set $\hbar \rightarrow 1$, 
but a dimensionful $\hbar$ can at any stage be easily restored by the
simple replacements $\partial_x \rightarrow \hbar\,\partial_x$ 
and $G^< \rightarrow \hbar^{-1}\,G^<$.

In the EWBG problem the system is both stationary and has planar 
symmetry along spacelike hypersurfaces of constant $z$ in the wall 
rest frame.  This means that in the plasma rest frame the mass has 
the form
\beq
m_{R,I} =  m_{R,I}(\gamma_w(z - v_w t)),
\eeq
where $v_w$ is the velocity of the phase transition front and 
$\gamma_w$ the corresponding boost factor. It is easiest to perform 
the analysis in the wall rest frame where $m_{R,I} = m_{R,I}(z)$.
Moreover, because equation (\ref{G_less_eom}) is Lorentz-covariant, one 
can immediately write the equations of motion in the wall frame. 
We also assume a stationary solution for $G^<$, so that  
its functional dependence can be parametrized in the wall frame 
as $G^<(k_\mu;z)$.~\footnote{To what extent the assumption
of stationarity can be relaxed is discussed at the end of this 
section.} Keeping the gradient corrections still to arbitrary order 
the equation of motion (\ref{G_less_eom}) then becomes
\beq
    \left( \gamma^0 k_0
         - \gamma^3( k_z - \frac{i}{2}\partial_z )
         - \vec \gamma \cdot \vec k_\|
         - m_R(z-\frac{i}{2}\partial_{k_z}) 
         - i\gamma^5 m_I(z-\frac{i}{2}\partial_{k_z}) 
    \right) iG^< = 0 .
\label{G_less_eom2}
\eeq
Despite of these simplifications, equation (\ref{G_less_eom2}) still
appears quite formidable to solve. Indeed, the Wightman 
function $G^<$ contains 16 independent real functions and 
equation (\ref{G_less_eom2}) consists of 16 complex equations for 
these components\footnote{To get a flavour of how the complete set 
of equations looks like, we refer to Refs.~\cite{ShinRafelski,
ZhuangHeinz-II}.}.
However, when $\vec k_\parallel \rightarrow 0$, {\em i.e.\ } for 
particles that move orthogonally to the wall, equation 
(\ref{G_less_eom2}) reduces to the much simpler 1+1 dimensional 
one studied in~\cite{KPSW1}.  We can make use of those results 
in the general case studied here if we can find a transform that removes the 
$\vec \gamma\cdot \vec k_\|\,$-term from equation 
(\ref{G_less_eom2}). In the static case under consideration this is done by
the boost $\Lambda$ to the frame 
in which $\vec k_\parallel = 0$. Obviously the four momentum in the 
boosted frame becomes
\beq
  \vec {\tilde k}_\| = 0
 ,\qquad     \tilde k_z = k_z, \qquad
 \tilde k_0 = 
 {\rm sign[k_0]}({k_0^2-\vec k_\|^{\,2}})^{\frac 12}.
\eeq
We define the boost velocity $\vec v_\| = \vec k_\|/k_0$ and the 
gamma-factor $\gamma_\| = k_0/\tilde k_0$ for later reference. 
Finding the spinor representation of the boost $\Lambda$ requires a little 
more work, but it can be deduced from the fact that the desired operator 
$L(\Lambda)$ must commute with $\gamma^3$ and $\gamma^5$ and it 
must effect the transform
\beq
  L(\Lambda)\kdag L^{-1}(\Lambda) \,=\,
     \gamma^0\tilde k_0 - \gamma^3 k_z \,\equiv \, \tilde{\kdag}.
\label{transfercond}
\eeq
The commutation requirements immediately suggest the form 
$L(\Lambda) = a + b \gamma^0 \vec \gamma \cdot \vec v_\|$, and
imposing the condition (\ref{transfercond}) then defines the 
coefficients $a$ and $b$, leading to the operator
\beq
  L(\Lambda) = \frac{k_0 + \tilde{k}_0 
               - \gamma^0\vec{\gamma}\cdot\vec{k}_{\|}}
                {\sqrt{2\tilde{k}_0(k_0+\tilde{k}_0)}}.
\label{L-Lambda}
\eeq
The inverse boost operator is simply 
$L^{-1}(\Lambda(k_0,\vec{k})) = L(\Lambda(k_0,-\vec{k}))$.
After performing the boost, we finally arrive at the following 
effectively 1+1-dimensional equation
\beq
    \left( \gamma^0 {\tilde k}_0
         - \gamma^3( k_z - \frac{i}{2}\partial_z )
         - m_R(z-\frac{i}{2}\partial_{k_z}) 
         - i\gamma^5 m_I(z-\frac{i}{2}\partial_{k_z}) 
    \right) i{\tilde G}^< = 0,
\label{G-less-eom3}
\eeq
where the boosted Wigner function $\tilde G^<$ is related to
the original one by
\beq
       \tilde G^< = L(\Lambda) G^<L(\Lambda)^{-1}.
\label{Gtilde}
\eeq

The further analysis of equation (\ref{G-less-eom3}) is identical 
to our earlier treatment~\cite{KPSW1}. However, for completeness  
we review the main physical aspects of the  derivation
here. The 
most important property of the 1+1-dimensional frame is that the 
differential operator in~(\ref{G-less-eom3}) is entirely spanned by the 
1+1-dimensional Clifford algebra, and that it commutes with the operator 
\beq
  \tilde{\!S}_z = \gamma^0\gamma^3\gamma^5,
\label{tildeSz}
\eeq
which measures spin $s$ in $z$-direction in which the wall moves. 
Therefore $s$ is a good quantum number in the boosted frame, and as a result 
we can restrict ourselves to finding solutions for $\tilde G^<$ that 
satisfy $\tilde{\!S}_z\tilde G^<_s = \tilde G^<_s \tilde{\!S}_z = 
s\tilde G^<_s$.  This condition immediately leads to the following 
decomposition for $\tilde G^<$, which is block-diagonal:
\beq
    - i\gamma^0\tilde{G}^{<}_s 
    = \frac{1}{4} 
   (\mathbf{1}+s\sigma^3) \otimes \rho^a\tilde{g}^s_a.
\label{Dec1+1}
\eeq
Here we made use of the chiral representation for the gamma matrices:
$\gamma^0 \; \rightarrow \; \rho^1$, 
$-i\gamma^0\gamma^5 \; \rightarrow \; \rho^2$,  
$-\gamma^5 \; \rightarrow \; \rho^3$ and
$\tilde{\!S}_z\equiv \gamma^0\gamma^3\gamma^5\; \rightarrow \;\sigma^3$,
where $\rho^a= ({\mathbf 1},\rho^i)$, and $\sigma^3$ and $\rho^i$ 
are the Pauli matrices. 
The normalization was chosen such that the component $\tilde g_0^s$ 
corresponds to the phase space density of states of spin $s$ in the 
boosted frame. All of the results derived in~\cite{KPSW1} apply for the 
functions $\tilde g_\alpha^s$ of course. From these we only need the 
constraint equations (here written to first order in gradients)
\beqa
&& k_0 \tilde g^s_0 - sk_z \tilde g^s_3 
       - m_R \tilde g^s_1 - m_I \tilde g^s_2 = 0
\label{CEA} \\
&& k_0 \tilde g^s_3 - sk_z \tilde g^s_0 
       + \frac12 m_R' \partial_{k_z} \tilde g^s_2
       - \frac12 m_I' \partial_{k_z} \tilde g^s_1 = 0
\label{CEB} \\
&& k_0 \tilde g^s_1 + \frac{s}{2} \partial_z \tilde g^s_2
       - m_R \tilde g^s_0 + \frac12 m_I' \partial_{k_z} \tilde g^s_3 = 0
\label{CEC} \\
&& k_0 \tilde g^s_2 - \frac{s}{2} \partial_z \tilde g^s_1
       - m_I \tilde g^s_0 - \frac12 m_R' \partial_{k_z} \tilde g^s_3 = 0.
\label{CED}
\eeqa
Equations~(\ref{CEB}-\ref{CED}) allow us to express the functions
$\tilde g^s_i$ in terms of $\tilde g^s_0$ alone, while the remaining
equation~(\ref{CEA}) contains the spectral condition for $\tilde g_0^s$.\\

We now make an explicit connection between the solutions in
the boosted and unboosted frame. To this end it is important to note that 
in the stationary case under study the spin in $z$-direction remains a 
good quantum number also in the unboosted frame. This is mathematically 
expressed by the fact that the spin operator in the original frame
\begin{equation}
  S_z  \equiv L^{-1}(\Lambda) \,\tilde{\!S}_z L(\Lambda)
  \; = \;  
 \gamma_\| \Big( \tilde S_z 
      - i(\vec v_\|\times \vec \alpha)_z \Big) 
\label{Sz}
\end{equation}
commutes with the differential operator in equation~(\ref{G_less_eom2}).
As a consequence, even in the original frame, the problem splits 
into two non-mixing sectors labeled by the spin $s$. 
By construction the Wightman function

\beq
  G^<_s \equiv L(\Lambda)^{-1} \tilde G^<_s L(\Lambda)
\label{Gtilde2}
\eeq
in the original frame commutes with the spin operator $S_z$.
Note that the spin $\vec S$ and $\vec\alpha = \gamma^0\vec\gamma$
transform under boosts as the magnetic and electric components of
an antisymmetric tensor, respectively. This can be seen from the
relations $\alpha^i = -i \sigma^{0i}$ and
$\tilde S_i = (1/2)\epsilon_{ijk} \sigma^{jk}$, where
$\sigma^{\mu\nu} = (i/2) [\gamma^\mu,\gamma^\nu]$ is the totally
antisymmetric tensor.

Now, the most general decomposition for the Wightman function $G^<_s$ 
in the unboosted frame can be written in the chiral representation
as~\cite{ElzeGyulassyVasak:1986,Bialynicki-BirulaGornickiRafelski:1991,
ShinRafelski}
\beq
    - i\gamma^0 G^<_s = \frac{1}{4} 
     \sigma^a \otimes \rho^b \, g^s_{ab},
\label{Dec3+1}
\eeq
where $\rho^a = (\mathbf{1},\rho^i)$ and 
$\sigma^a = (\mathbf{1},\sigma^i)$
\footnote{It is easy to show that $\rho^a \otimes \sigma^b$ span the 
full 3+1 dimensional Clifford algebra.}. The connection between
the component functions in~(\ref{Dec3+1}) and (\ref{Dec1+1})
is now  found\footnote{An alternative way to derive this connection 
would be to identify the transformation properties of $g_{ab}^s$ by 
considering the fermionic currents, see section (\ref{sec:currents}).} 
by direct use of the relation (\ref{Gtilde}) and the explicit form 
of the boost operator (\ref{L-Lambda}):
\def\ps{{\phantom{-}s}}
\beq
   g^s_{ab} = \left( \begin{array}{cccc}
  \gvp  \tgs_0 &   \tgs_1       &    \tgs_2       & \gvp\tgs_3  \\ 
  \vgva \tgs_3 &\ps \vgvb \tgs_2& -s \vgvb\tgs_1  & \vgva\tgs_0 \\
  \vgvb \tgs_3 & -s \vgva \tgs_2& \ps \vgva\tgs_1 & \vgvb\tgs_0 \\
  s \tgs_0     & s \gvp \tgs_1  &  s \gvp\tgs_2   & s \tgs_3
   \end{array}\right).
\label{connect}
\eeq
\vskip4truemm

Given the transformation (\ref{connect}) and the constraint 
equations (\ref{CEA}-\ref{CED}), we are now ready to deal with 
the equations of motion (\ref{G_less_eom2}). Out of the 16 
equations we are primarily interested in the 00-component, 
defining the kinetic properties of $g^s_{00}$, which corresponds 
to the particle number density in the phase space.
Other components can always be determined from $g^s_{00}$ by the 
use of constraint equations, and their physical significance will 
be discussed in the next section. The relevant kinetic equation 
has the form
\beq
%      \del_t g^s_{00} 
%    + \del_i g^s_{i3} 
      \del_z g^s_{33} 
    - m_R'\del_{k_z} g^s_{01}
    - m_I'\del_{k_z} g^s_{02} = 0
\label{g00_kinetic}
\eeq
and the constraint equation reads
\beq
   k_0 g^s_{00} - k_i  g^s_{i3} - m_R g^s_{01} - m_I g^s_{02} = 0.
\label{g00_constraint}
\eeq
In (\ref{g00_kinetic}) we have kept terms to the second and in 
(\ref{g00_constraint}) to the first order in gradients, which 
suffices for the derivation of the kinetic and constraint equations 
accurate to order $\hbar$. Using (\ref{connect}) and the constraints 
(\ref{CEA}-\ref{CED}) it is now a simple matter to show that we 
get the following kinetic equation
\beq
   k_z \del_z g^s_{00}
 - \left(\frac{1}{2}{|m|^2}'\delk 
 - \frac{s}{2\tilde{k}_0}(|m|^2\theta')'\delk
    \right) g^s_{00} = 0,
\label{kinetic3+1}
\eeq
and the {\em algebraic} constraint equation
\beq
\left(k^2 - |m|^2 + \frac{s}{\tilde{k}_0}(|m|^2\theta') 
    \right) g^s_{00} = 0 ,
\label{constraint3+1}
\eeq
which both contain nontrivial CP-violating terms $\propto \theta'$,
where the angle $\theta$ corresponds to the complex phase of the mass 
term $m = |m|(z)e^{i\theta(z)}$. 

The most important outcome of these results follows from the 
observation that the equation~(\ref{constraint3+1}) has a 
spectral solution\footnote{The normalization of $\tilde{G}^{<}$
in~(\ref{Dec1+1}), combined with the spectral sum rule
$i\gamma^0\int dk_0 (G^>-G^<) = 2\pi \mathbf{1}$, fixes the 
normalization in (\ref{spectral-dec}) unambiguously. It differs
by a factor of 4 from our definition in~\cite{KPSW1}, which was 
thus not entirely consistent.}
\beqa
g^{s}_{00}  &\equiv& \sum_\pm \frac{2\pi }{Z_{s\pm}}
                 \, n_s \,\delta(k_0 \mp \omega_{s\pm}),
\label{spectral-dec}
\eeqa
where $\omega_{s\pm}$ denotes the dispersion relation
\beq
     \omega_{s\pm} = \omega_0 \mp s \frac{|m|^2\theta'}
   {2\omega_0({\omega_0^2-\vec k_{\parallel}^{\,2}})^{1/2}},
     \qquad \qquad \omega_0 = \sqrt{ \vec k^2 + |m|^2}
\label{dispersion1}
\eeq
and $Z_{s\pm} = 
  1 \mp s|m|^2\theta'/2({\omega_0^2-\vec k_{\parallel}^{\,2}})^{3/2}
%{\omega_{s\pm}}/{\omega_0}
$. Because of the delta 
function in (\ref{spectral-dec}) the functions $n_s(k_0,k_z,z)$ are 
projected on-shell and become the distribution functions $f_{s+}$ 
and $f_{s-}$ for particles and antiparticles with spin $s$, 
respectively, defined by 
\beqa
     f_{s+} &\equiv& n_s(\omega_{s+},k_z,z) \nonumber
 \\
     f_{s-} &\equiv& 1 - n_s(-\omega_{s-},-k_z,z).
\label{fs}
\eeqa
This on-shell projection proves the implicit
assumption underlying the semiclassical WKB-methods, that the plasma
can be described as a collection of single-particle excitations with
a nontrivial space-dependent dispersion relation. 

Integrating (\ref{kinetic3+1}) over the positive and negative 
frequencies and taking account of~(\ref{spectral-dec}) and~(\ref{fs})
we now get the following on-shell kinetic equations:
\begin{equation}
%        \del_t f_{s\pm} +
       v_{s\pm} \del_z f_{s\pm}
      + F_{s\pm} \partial_{k_z} f_{s\pm} = 0,
\label{ke-fspm0}
\end{equation}
where the quasiparticle {\it group velocity} $v_{s\pm} \equiv 
k_z/\omega_{s\pm}$ is expressed in terms of the kinetic momentum 
$k_z$ and the quasiparticle energy $\omega_{s\pm}$~(\ref{dispersion1}), 
and the {\em semiclassical force} 
\begin{eqnarray}
    F_{s\pm} = - \frac{{|m|^2}^{\,\prime} }{2\omega_{s\pm}}
                   \pm  \frac{s(|m|^2\theta^{\,\prime})^{\,\prime}}
                   {2\omega_0({\omega_0^2-\vec k_\parallel^{\,2}})^\frac 12}.
\label{Fspm}
\end{eqnarray}
When compared with the 1+1 dimensional case the sole, but significant,
 difference in the force is that the CP-violating 
$\theta'$-term is enhanced by the boost-factor $\gamma_\parallel 
= \omega_0/({\omega_0^2-\vec k_\parallel^{\,2}})^\frac 12$. 

\vskip4truemm

So far we have considered the Boltzmann equation~(\ref{ke-fspm0})
for a static situation (in the wall frame) with planar symmetry. It is 
interesting to see to what extent these restrictions can be relaxed.
The essential aspect of our treatment above 
%\footnote{Note that the commutation condition~(\ref{spin-commutation})
%is equivalent to requiring ${\cal D} G^<_s = 0$ to give
%a consistent set of equations for the components
%$\tilde g_a^s$ $(a=0..3)$.
was the requirement of spin conservation in $z$-direction,
guaranteeing the consistency of the spin conserving decomposition
$G^<\rightarrow G^<_s$ of the Wigner function. This requirement 
can be formally expressed by the condition
\beq
  [S_z,{\cal D}]G^<_s = 0
,
\label{comm-cond}
\eeq
where ${\cal D}$ is the derivative operator in~(\ref{G_less_eom}), and
the commutator evaluates to
\begin{equation}
[\vec S_z,{\cal D}] = 
  - \gamma_\| \gamma^0 \vec v_\| \times \partial_{\vec x_\|}
  +  \gamma_\|
 \vec \gamma_\| \times (\partial_{\vec x_\|} + \vec v_\|\,\partial_t)
.
\label{spin-commutation}
\end{equation}
The condition~(\ref{comm-cond}) 
is clearly satisfied when 
the parameter dependencies of the Wigner function are of the form
\footnote{
We have checked that, when~(\ref{transients-form})
is inserted in ${\cal D} G^<_s = 0 $, one obtains a fully consistent
set of 16 equations for the components $\tilde g_a^s$ $(a=0..3)$.
}
\begin{equation}
     G_s^< = G_s^< (k_\mu;z,t-\vec v_\|\cdot \vec x_\|).
\label{transients-form}
\end{equation}
For static, planar symmetric solutions studied so far 
the commutation condition is satisfied trivially, 
because for them $\partial_t G^< = \partial_{x_\|}G^< = 0$. 
We note that in the $1+1$-frame (in which $\vec k_\| = 0$) 
our result~(\ref{transients-form}) is particularly simple and intuitive. 
In this frame $\partial_{\tilde t}$ is spin conserving, while 
$\tilde \partial_{\vec x_\|}$ violates spin, and thus
the only off-diagonal contributions in the equation
$\tilde{\cal D} \tilde G^<_s = 0$ are the 
$\tilde \partial_{\vec x_\|}$-derivatives, leading to the requirement 
$\tilde \partial_{\vec x_\|} \tilde g_{a}^s = 0$.
This is completely consistent with~(\ref{transients-form}), which in 
the 1+1-frame reduces to  
$\tilde G^<_s = \tilde G^<_s (\tilde k_\mu;z,\tilde t\,)$.

Following the procedure outlined earlier in this section
one can show that there are no new corrections to the constraint 
equation~(\ref{constraint3+1}), implying that the 
transients~(\ref{transients-form}) respect the simple algebraic on-shell
condition. The kinetic equation~(\ref{ke-fspm0}) now generalises to 
\begin{equation}
       \del_t f_{s\pm}
      + \vec v_{s\pm} \cdot \del_{\vec x} f_{s\pm}
      + F_{s\pm} \partial_{k_z} f_{s\pm} = 0,
\label{ke-fspm}
\end{equation}
where $\vec v_{s\pm} \equiv \vec k/\omega_{s\pm}$, and the
distribution function satisfies 
$f_{s\pm} = f_{s\pm}(k_\mu; z, \, t - \vec v_\| \cdot \vec x_\|)$,
where now $\vec v_\| \equiv \vec k_\|/\omega_{s\pm}$. 
We shall assume this to be satisfied when discussing 
the current continuity equations in the next section.

\section{Fermionic currents and CP-violating sources}
\label{sec:currents}

In order to shed light on the physical significance of the various
components of the matrix $G^<_s$, it is instructive to consider 
the fermionic currents 
\beqa
&&  \left< \bar{\psi}(x) \Gamma \psi(x) \right> 
    = - \int \frac{d^4k}{(2\pi)^4} 
    \mbox{Tr}\left[ \Gamma iG^<(x,k)\right] \\
&&  \Gamma = \left( 
     \mathbf{1} , \gamma^5 , \gamma^\mu , \gamma^5\gamma^\mu , 
         \sigma^{\mu\nu}
     = \frac{i}{2}[\gamma^\mu,\gamma^\nu]
             \right) .
\eeqa
By direct computation one sees that $g^s_{01}$ and  $g^s_{02}$ 
correspond to a scalar and pseudoscalar density, respectively, while 
$(g^s_{00}, g^s_{i3})$ forms a 4-vector, and $(g^s_{03}, g^s_{i0})$ a 
4-pseudovector.  Finally, $g^s_{i1}$ and $g^s_{i2}$ are components of 
an antisymmetric tensor, and so they transform like the magnetic
and electric field, respectively. These transformation properties 
could actually have been used as an alternative and elegant way
to infer the relation between the boosted and nonboosted frames in
the previous section. 

 Let us now look at the vector and pseudovector
currents in a little more detail. By writing $g^s_{ab}$ in terms of 
$g^s_{00}$ and dropping the total derivative terms we find 
\beqa
j_s^\mu   &=& \int \frac{d^4k}{(2\pi )^4} 
  (1 ; \vec v) g^s_{00} \nonumber \\
j^\mu_{5s} &=& \int \frac{d^4k}{(2\pi )^4} 
  (1 ; \vec v_\|, \frac{1}{\gamma_\|^2v_s})sv_s\gvp g^s_{00}.
\eeqa
These expressions are clearly the expected generalizations with
appropriate boost factors of the 1+1 dimensional currents 
derived in~\cite{KPSW1}.

It is straightforward to show from the equations of motion, that
the vector current divergence is conserved, and the axial current
has the usual form (since we do not treat the gauge fields here, 
the axial current is taken to be nonanomalous;
for an account of the anomaly see~\cite{Kainulainen:2002ah}):
\beqa
\partial_\mu j_s^\mu &=& 0 
\label{vecdiv1}
        \\
\partial_\mu j^\mu_{5s} &=& 
           2im_R \langle\bar{\psi_s}\gamma^5\psi_s \rangle
           - 2m_I \langle\bar{\psi_s} \psi_s \rangle.
\label{axdiv1}
\eeqa
If we define the fluid density\footnote{$n_{s\pm}$ should not be
confused with the phase space density $n_s$ in~(\ref{spectral-dec}).}
$n_{s\pm}$ and fluid velocity moments as\footnote{Here we defined 
integration over the positive frequencies. To obtain equations for 
antiparticles one integrates over negative frequencies taking into 
account the appropriate thermal equilibrium limit in (\ref{fs}).}
\begin{eqnarray}
  n_{s+}  &\equiv& \int_+\frac{d^{\,4}k}{(2\pi)^4} g^s_{00}
\nonumber\\
  \langle v_{s+}^p \rangle &\equiv& 
        \frac{1}{n_{s+}} \int_{+}\frac{d^{\,4}k}{(2\pi)^4}
        \left( \frac{k_z}{k_0} \right)^p g^s_{00} \,,
\label{Ncont-2}
\end{eqnarray}
then Eqs.~(\ref{vecdiv1}-\ref{axdiv1}) can be shown to correspond
to the two lowest velocity moments of the Boltzmann equation
(\ref{ke-fspm}):
\beqa
   \del_t n_{s\pm}
    + \partial_z(n_{s\pm} \langle v_{s\pm}\rangle) &=& 0,  
\label{Ncont1}
\\
   \partial_t \left(n_{s\pm} \langle v_{s\pm} \rangle  \right)
     + \del_z \left(n_{s\pm} \langle v_{s\pm}^2 \rangle \right)
     &=&  S_{s\pm}.
\label{Ncont2}
\eeqa
The quantity $S_{s\pm}$ is related to the baryogenesis 
source and is given by the average over the semiclassical force 
(\ref{Fspm}) divided by $k_0$:
\beq
    S_{s\pm} =
    - \frac 12 \vert m \vert^{2\,\prime} {\cal I}_{2s\pm}
    \pm \frac 12  \, s(\vert m \vert^{2}\theta')' \,  
      {\cal I}_{3s\pm},
\label{realsource}
\eeq
and
\beq
{\cal I}_{ps+} \equiv 
 \int_{+}\frac{d^4 k}{(2\pi)^4} \frac{g_{00}^s}{k_0^p}.
\label{iota}
\eeq

It should be stressed that while equations (\ref{Ncont1}-\ref{Ncont2}) 
are illustrative in relating the baryogenesis source in moment equations 
to the non-conservation of the axial current, they are somewhat formal, 
and not directly suitable for actual calculations. 
For practical baryogenesis calculations it is convenient to treat 
the Boltzmann equation (\ref{ke-fspm}) in linear response approximation
with respect to deviations from thermal equilibrium 
\begin{equation}
 \delta f_{s\pm} \equiv f_{s\pm} - f_{0s\pm}
,\qquad
f_{0 s\pm}  = \frac{1}{e^{\beta \gamma_w(\omega_{s\pm} + v_w k_z)}+1},
\label{distribution-fn-eq}
\end{equation}
where $f_{0 s\pm}$ is the equilibrium distribution function in the 
wall frame, $\beta =1/T$, $v_w$ is the wall velocity and 
$\gamma_w$
%= 1/\sqrt{1-v_w^2}$. 
is the corresponding boost factor. Working to linear order in $v_w$ 
we can rewrite~(\ref{ke-fspm}) as 
\begin{equation}
    \left(\del_t  + \vec v_{s\pm}\cdot\del_{\vec x} 
          + F_{s\pm} \partial_{k_z}\right) \delta f_{s\pm} 
   + v_w F_{s\pm} (\partial_{\omega} f_{\omega})_{\omega=\omega_{s\pm}} 
   = 0,
\label{boltzeqn-split}
\end{equation}
with $f_{\omega} \equiv 1/(e^{\beta\omega}+1)$. 
Equation~(\ref{boltzeqn-split}) implies the following kinetic equation 
for the CP-violating density 
$\delta f_{s} \equiv \delta f_{s+}-\delta f_{s-}$:
\begin{equation}
    \left(\del_t  + \frac{\vec k}{\omega_{0}}\cdot \del_{\vec x} 
  - \frac{{|m|^2}'}{2\omega_{0}} \partial_{k_z}\right) \delta f_{s} 
  = -  v_w \left( \delta F_{s} (\partial_{\omega} f_{\omega})_{\omega_{0}} 
     +  F_{0}\delta\omega_{s}
 \bigg[\Big(\frac{\partial_{\omega} f_{\omega}}{\omega}\Big)_{\omega_{0}}
      - (\partial^2_{\omega} f_{\omega})_{\omega_{0}} \bigg] \right),
%   = 0,
\label{boltzeqn-split-2}
\end{equation}
where we kept only the leading second order terms in gradients, and 
\begin{eqnarray}
    F_{0} &\equiv& - \frac{{|m|^2}'}{2\omega_{0}}
\nonumber\\
    \delta\omega_{s} 
           &\equiv& s\frac{|m|^2\theta'}{\omega_{0}\tilde\omega_{0}}
\nonumber\\
\delta F_{s} &\equiv& F_{s+} - F_{s-} 
              = s\frac{(|m|^2\theta')'}{\omega_{0}\tilde\omega_{0}},
\label{boltzeqn-split-3}
\end{eqnarray}
where $\tilde\omega_0 = ({\omega_0^2-\vec k_\|^2})^{1/2}$.
Equation~(\ref{boltzeqn-split-2}) displays two very important features
not evident in equations (\ref{Ncont1}-\ref{Ncont2}): the CP-violating
perturbation $\delta f_{s}$ 
is only sourced by terms second order in derivatives (first order in 
$\hbar$) {\em and} the source is suppressed by the wall velocity.  

To get equations suitable for baryogenesis calculations it would be 
necessary to include collision terms in~(\ref{boltzeqn-split-2}), 
which are responsible for all thermalisation and transport processes. 
While treating collisions is beyond the scope of this paper, 
let us nevertheless mention that an approximate solution for 
Eq.~(\ref{boltzeqn-split-2}) with collisions is typically found 
by taking moments of the kinetic equation, leading to a set of fluid 
equations for the moments. In order to get a good approximation while 
using only a few lowest moments, it is important to parametrize 
$\delta f_{s}$ such that the parametrization models the actual solution in 
the best possible way. Particularly suitable and often used is the fluid 
ansatz parametrised by the chemical potential and a perturbation 
that is an odd function of the momentum. The chemical potential then 
captures an excess of particles formed at the phase boundary front, 
while the latter parametrizes the anisotropic bulk motion of the fluid. 
For a more detailed account of the fluid equations we refer the reader 
to Refs.~\cite{JPT-thick,MooreProkopec95,ClineJoyceKainulainen-II,
HuberJohnSchmidt}.

 We finally note that the continuity equations for the vector and axial
vector current~(\ref{Ncont1}-\ref{Ncont2}) are equivalent to the two
lowest order moments of the Boltzmann
equation~(\ref{boltzeqn-split-2})~\cite{KPSW1}, and hence also 
to the fluid equations.

%%%%%%%%%%%%%%%%%%%%%%%%%%%%%%%%%%%%%%%%%%%%%%%%%%%%%%%%%%%%
%
% WKB METHOD
%
%%%%%%%%%%%%%%%%%%%%%%%%%%%%%%%%%%%%%%%%%%%%%%%%%%%%%%%%%%%%
%
\section{WKB method}
\label{WKB method}

For the sake of comparison we shortly review the 
derivation of the semiclassical transport equations via the
WKB-method~\cite{JPT-letter,JPT-thick,ClineJoyceKainulainen-II},
in which the dispersion relations of the single particle excitations
are derived directly from the Dirac equation (\ref{Diraceqn}) in 
the gradient expansion.  In the wall frame the mass~(\ref{mass1}) 
is only a function of $z$, $m(z)=|m(z)|e^{i\theta(z)}$, and 
therefore the energy and momentum parallel to the wall are 
conserved quantities. We can then seek the solutions of the form 
$\psi(x) = \exp(-ik_0t+i\vec k_\| \cdot \vec x_{\|})
\psi_{k_0\vec k_\parallel}(z)$, which gives
\beq
\left(\gamma^0k_0 - \vec \gamma\cdot \vec k_\parallel
- i  \gamma^3 {\del}_z - m_R - i\gamma^5 m_I \right) 
  \psi_{k_0\vec k_\parallel}(z) = 0.
\label{wkb.1}
\eeq
As in the previous chapter, we can use the Lorentz boost $\Lambda$
(\ref{L-Lambda}) to get rid of the $\vec\gamma \cdot \vec k_\parallel$ 
term in (\ref{wkb.1}).  The remaining part of the Dirac operator then 
commutes with the spin operator $\tilde S_z$ (\ref{tildeSz}), indicating 
that the spin $s$ in $z$-direction is a good quantum number, and one can
write the boosted wave function 
$\tilde\psi_{\tilde k_0} = L(\Lambda)\psi_{k_0k_\parallel}$
as a direct product of $2\times 2$ spinors, 
\beq
\tilde\psi_{\tilde k_0}\equiv\tilde\psi^s_{\tilde k_0}\otimes \chi_s,
\qquad \tilde\psi^s_{\tilde k_0} = 
               \left(\begin{array}{c}
                     L_s \cr
                     R_s 
                      \end{array}
               \right),
\label{wkb.3}
\eeq

where $\chi_s$ is the eigenspinor of the spin operator~(\ref{tildeSz})
in the $2\times 2$ spinor space, $\sigma^3 \chi_s = s\chi_s$, and
$\tilde k_0 = {\rm sign}[k_0]( k_0^2-k^2_{||})^{1/2}$ as usual. With 
this decomposition Eq.~(\ref{wkb.1}) simplifies to the following two 
equations for the left and right chiral states in the boosted frame:
\beqa
 (\tilde k_0 - i s\partial_z) L_s &=& m R_s
\nonumber\\
  (\tilde k_0 + i s\partial_z) R_s &=& m^* L_s.
\label{wkb.4}
\eeqa
Our analysis so far differs from the ones in the literature 
in that we have discussed more explicitly the boost and its relation 
to the spin-decomposition. The following steps in the analysis 
are quite standard, and we refer the reader for example 
to~\cite{ClineJoyceKainulainen-II,Kainulainen:2002ah} 
for more details. Solving $R_s$ in terms of $L_s$ in 
(\ref{wkb.4}) and then introducing the WKB-parametrization
\beq
 L_s = u_s e^{i\int^z k_s dz'},
\eeq
one can write down a set of two coupled second order equations for
$u_s$ and $k_s$. Solving these equations to second order in gradients
gives the dispersion relation
\beq
  k_s =  p_0   +  \scp \frac{(s k_0 + p_0)\theta'}{2p_0} ,
\label{wkb.10}
\eeq
where $p_0 = {\rm sign}(k_s)\sqrt{\tilde k_0^2 - |m|^2}$ and 
$\scp = 1 (-1)$ for particles (antiparticles). Remembering 
that\footnote{The transform back from $\tilde k_0$ to $k_0$ 
was not effected in the derivation presented 
in~\cite{ClineJoyceKainulainen-II}, leading to a missing 
$\gvp$-factor in the CP-violating force. The gamma-factor 
was included in~\cite{CJK_erratum} however.}
$\tilde k_0^2 = k_0^2 - k_\|^2$ and inverting (\ref{wkb.10}) 
one can write the energy in the unboosted frame as
\beq
  k_0 \simeq  
   \sqrt{(k_s - \scp \frac{s\theta'}{2})^2 + k_\|^2 + |m|^2}
              - \scp \frac{s\theta'}{2}.
\label{wkb.11}
\eeq
From this expression for the energy we can finally compute
the physical momentum, which is defined in terms of the group 
velocity $v_g = k_0 (\partial_{k_s}k_0)_z$:
\beq
  k_z \equiv k_0 v_g
  = p_0   +  \scp \frac{s|m|^2\theta'}{2p_0\tilde k_0}
\label{wkb.12}
\eeq
and the corresponding physical force 
$F_z = \dot k_z =\omega_{s\pm} dv_{s\pm}/dt$:
\beq
F_z  =  - \frac{{|m|^2}'}{2k_0}
        + \scp \, \frac{s(|m|^2\theta')'}{2k_0\tilde k_0}.
\label{voima}
\eeq
These results are in complete agreement with those found in 
section~2, completing the explicit proof of correctness 
of the semiclassical limit in the present paper.

Let us finally point out that the kinetic equations could
be formulated in terms of the canonical 
momentum (\ref{wkb.10}), as it was done originally 
in~\cite{JPT-thick,ClineJoyceKainulainen}. 
The transformation that relates the kinetic and canonical momenta is
noncanonical:
\begin{equation}
   k_z = k_s \left( 1 \mp \frac{s\theta'}{2(k_s^2 + |m|^2)^{1/2}}\right).
\label{wkb.13}
\end{equation}
The advantage of the canonical formulation is that it allows for a 
hamiltonian formulation of kinetic equations, while the disadvantage
is that the canonical momentum and energy do not transform as a 4-vector
under Lorentz boosts. This makes the formulation of fluid equations
cumbersome, and so far it has not been done with due care in the
literature~\cite{JPT-letter,JPT-thick,ClineJoyceKainulainen,HuberJohnSchmidt}.

%%%%%%%%%%%%%%%%%%%%%%%%%%%%%%%%%%%%%%%%%%%%%%%%%%%%%%%%%%%%
%
% APPLICATIONS TO SUPERSYMMETRIC MODELS
%
%%%%%%%%%%%%%%%%%%%%%%%%%%%%%%%%%%%%%%%%%%%%%%%%%%%%%%%%%%%%
%
\section{Applications to supersymmetric models}
\label{Applications to supersymmetric models}

When studying electroweak baryogenesis in supersymmetric models
an important source comes from CP-violating dynamics of charginos
in presence of a bubble wall. In this case a formalism that describes 
the dynamics of mixing fermions is required. Introducing the mixing 
does not bring any extra complications {\em w.r.t.}~to extension
to 3+1 dimensions and hence, proceeding just as in 
KPSW-I~\cite{KPSW1}, we find the semiclassical Boltzmann equation
\begin{equation}
        \del_t f_{si\pm}
      + \vec v_{si\pm} \cdot \del_{\vec x} f_{si\pm}
      + F_{si\pm} \partial_{k_z} f_{si\pm} = 0 \,,
\label{boltzeqn-2}
\end{equation}
where $f_{si+}$ and $f_{si-}$ are the distribution functions for 
fermions and antifermions of spin $s$, respectively, written in 
the propagating basis in which the mass matrix is diagonal. As 
it was the case in Eq.~(\ref{ke-fspm}), $f_{si-}$'s are assumed
to satisfy the condition $f_{si\pm} = 
f_{si\pm}(k_\mu; z, t - \vec v_\| \cdot \vec x_\|)$.

The group velocity of the $i$'th quasiparticle eigenstate equals 
$\vec v_{si\pm} = \vec k/\omega_{si\pm}$ and the CP-violating 
semiclassical force reads
\beq
F_{si\pm}  =  - \frac{{|M_{i}|^2}' }{2\omega_{si\pm}}
              \pm \frac{s(|M_{i}|^2\Theta_i')'}
   {2{\omega_{0i}}({\omega_{0i}^2-\vec k_{\parallel}^{\,2}})^{1/2}},
\label{scforce2}
\eeq
where $|M_i|^2 \Theta_i'$ is a diagonal matrix which can be computed
from the original mass matrix and the rotation matrix $U$ diagonalizing
$MM^\dagger$ as follows:
\beq
     |M_d|^2 \Theta'
            = \mbox{Im}\Big( U M{M'}^\dagger U^\dagger \Big)_d .
\label{trans1}
\eeq
For example, in the case of charginos the mass term reads
\begin{equation}
{\overline\Psi_R}\, M\, \Psi_L + {\rm h.c.}
\,,
\label{chargino-mass}
\end{equation}
where $\Psi_R = ({\tilde W}_R^+,{\tilde h}^+_{1,R})^T$
and $\Psi_L = ({\tilde W}_L^+,{\tilde h}^+_{2,L})^T$
are chiral fields consisting  of winos and higgsinos.
The mass matrix is conveniently parametrized as 
\begin{equation}
        M = \left( \begin{array}{cc} m_2  & gH_2^* \\
                                     gH_1^* & \mu     
                   \end{array} \right)\,,
\label{CharginoMatrix}
\end{equation}
where $H_1=h_1 e^{i\theta_1}$ and $H_2=h_2 e^{i\theta_2}$ are the 
Higgs field vacuum expectation values and $\mu$ and $m_2$ are the 
soft supersymmetry breaking parameters. For a realistic choice of 
parameters there is no spontaneous CP-violation in the MSSM, so to 
good approximation we can take the Higgs {\it vev's} to be 
real~\cite{HuberJohnLaineSchmidt,HuberJohnSchmidt-I}. In this
case~\cite{ClineJoyceKainulainen,ClineJoyceKainulainen-II,KPSW1}
\beq
 |M_i|^2 \Theta_i' \Rightarrow   m^2_\pm \Theta_\pm'  =
         \mp  \frac{g^2}{\Lambda}\Im (\mu m_2) ( h_1h_2 )'
\,,
\label{chargino-source}
\eeq
where $\Theta_+$ ($\Theta_-$)  corresponds to the higgsino-like 
state when $|\mu| > |m_2|$ ($|\mu| < |m_2|$), and 
$\Lambda = m_+^2-m_-^2$ is the mass splitting of the chargino mass 
eigenstates~\cite{ClineJoyceKainulainen,ClineJoyceKainulainen-II,KPSW1}. 
CP violation is here mediated via the parameters $\mu$, $m_2$ and may 
be large~\cite{PilaftsisWagner}. Yet another interesting CP-violating 
source arises when the Higgs condensate develops a CP-violating 
phase, as it is in the case of the NMSSM~\cite{HuberJohnSchmidt}
(for details of computation see~\cite{KPSW1}).

%%%%%%%%%%%%%%%%%%%%%%%%%%%%%%%%%%%%%%%%%%%%%%%%%%%%%%%%%%%%
%
% CONCLUSIONS AND DISCUSSION
%
%%%%%%%%%%%%%%%%%%%%%%%%%%%%%%%%%%%%%%%%%%%%%%%%%%%%%%%%%%%%
%
\section{Conclusions and discussion}
\label{Conclusions and discussion}

We have presented a first principle derivation of the semiclassical 
limit of transport equations for a collisionless gas of fermions in a 
spatially varying background. 
Our analysis is based on the gradient expansion in the relative 
coordinate of the Kadanoff-Baym equations of 
motion~\cite{KadanoffBaym:1962} for the 
out-of-equilibrium two point function $G^<$. The most 
crucial step in our derivation was to show that to first order in 
$\hbar$ the Wigner transformed function $G^<$ has a {\em spectral 
solution}.  The transport equations for the corresponding on-shell 
excitations are shown to reduce to the usual Liouville form, and  
contain a CP-violating semiclassical force of order $\hbar$.
We also showed how the same force can be obtained by WKB 
methods~\cite{JPT-letter,JPT-thick,ClineJoyceKainulainen,
ClineJoyceKainulainen-II,HuberJohnSchmidt}.
The force is suitable for baryogenesis calculations based on the
charge transport picture~\cite{CohenKaplanNelson} at a first order 
electroweak phase transition.

The present work extends our earlier 1+1 dimensional 
results~\cite{KPSW1} to the more realistic case of a planar 
phase transition wall in 3+1 dimensions, relevant for EWBG.
The problem becomes 
analytically tractable only upon realising that the spin orthogonal 
to the moving planar wall remains a good quantum number for particles 
moving in arbitrary directions with respect to the wall. Thanks to this
conserved quantity the problem can be reduced to the dynamics of one 
spin dependent distribution function for fermions, and one for 
antifermions. We assumed planar symmetry and stationarity 
appropriate for EWBG. For completeness we also discussed 
how our formalism can be extended to include particular time and
space dependent transients which conserve spin.

The spectral solution for $G^<$ can only be found up to order $\hbar$.
Beyond order $\hbar$ the spectral decomposition ansatz is not consistent
and the transport equations do not have the semiclassical Liouville limit.
It is therefore very fortunate that in at least some of the most 
interesting cases~\cite{ThickWalls} the phase transition walls are
rather wide $\ell_{\rm wall} \gg \ell_{\rm dB}$, where $\ell_{\rm dB}$
is the de Broglie wave length $\sim 1/T$ of a particle in the plasma,
and so the semiclassical equation is expected to be a reasonable
approximation.

The case of mixing fermions is relevant for example in supersymmetric 
baryogenesis scenarios, in which baryogenesis can be mediated by 
mixing of charginos or neutralinos and other species. Supersymmetric
models, in contrast to the minimal version of the standard 
model~\cite{KajantieLaineRummukainenShaposhnikov}, are 
shown to be viable candidates for EWBG~\cite{CarenaQuirosWagnerETAL}.
We therefore derived explicitly the CP-violating force for a
general case of mixing fermions and for charginos in particular.
With this, the present work covers the proof of the semiclassical 
methods used for baryogenesis calculations in the literature so 
far~\cite{ClineJoyceKainulainen-II,HuberJohnSchmidt}. The directions 
for further work include the inclusion of the self-consistent
gauge (hyperelectric) fields~\cite{Kainulainen:2002ah} in 3+1 
dimensions and the study of interactions, which introduce new
collisional baryogenesis sources~\cite{Kainulainen:2002sw}.

\section*{Acknowledgements}

We wish to thank Michael Joyce for earlier collaboration on the
Schwinger-Keldysh formalism for fermions, and Dietrich B\"odeker
for raising a doubt about the validity of the spectral decomposition
ansatz as a general tool for the Kadanoff-Baym equations.

%%%%%%%%%%%%%%%%%%%%%%%%%%%%%%%%%%%%%%%%%%%%%%%%%%%%%%%%%%%%
%
% references
%
%%%%%%%%%%%%%%%%%%%%%%%%%%%%%%%%%%%%%%%%%%%%%%%%%%%%%%%%%%%%

\end{document}